# Unambiguous evidence of three coexisting ferroelectric phases in a lead-free Li$_x$Na$_{1-x}$NbO$_3$ system

Abhinav Kumar Singh,[1] Digvijay Nath Dubey,[1] Gurvinderjit Singh,[2] and Saurabh Tripathi[1, a)]
[1)] *Department of Physics, Indian Institute of Technology (BHU), Varanasi, 221005, India*
[2)] *Laser and Functional Materials Division, Raja Ramanna Centre for Advanced Technology, Indore, 452013, India*

(Dated: 12 May 2020)

We report here the presence of three coexisting ferroelectric phases in a lead-free Lithium Sodium Niobate (Li$_x$Na$_{1-x}$NbO$_3$; LNNx) system stable for $0.15 \leq x \leq 0.80$, which contrasts the review report of Dixon et al. [C. A. Dixon and P. Lightfoot, Physical Review B **97**, 224105 (2018)]. More importantly, we have identified LNN20 as an important composition for technological applications, due to its high dielectric permittivity, low loss, and high ferroelectric response. The anomalous dielectric and ferroelectric responses in LNN20 have been attributed to the MPB like nature around this composition.

Lead-based materials are known to have detrimental effects on health/environment for a long time, and therefore, international efforts are continuously increasing to restrict the use of such toxic materials from everyday life.[1,2] However, most of the piezodevices based industries still use lead zirconate titanate (PZT), due to its unrivaled high electromechanical response, which has inspired researchers to look for lead-free alternatives. Most of the research for such applications are focussed on the perovskite materials; BaTiO$_3$, NaNbO$_3$, LiNbO$_3$, KNbO$_3$, Na(Bi$_{0.5}$Ti$_{0.5}$)O$_3$, their solid solutions, and derivatives[3–9]. Alkali Niobates have drawn considerable interest after the discovery of high piezoelectric properties in Na$_x$K$_{1-x}$NbO$_3$[4,5]. Sodium Niobate (NN) and Lithium Niobate (LN) are some of the important materials falling under this category having several technological applications, such as high-density optical data storage, nonlinear photonics, electro-optic devices, SAW (surface acoustic wave) devices, laser modulators, piezoelectric devices, optical wave-guides etc[10–18]. NN has an anti-ferroelectric orthorhombic *Pbma* (O$_{AFE}$) structure at room temperature and shows a series of phase transitions as a function of temperature with complex structures, driven by in phase and out of phase rotation of oxygen octahedra[19–21]. On the other hand, LN has a ferroelectric rhombohedral *R3c* (R$_{Li}$) phase at room temperature, which transforms into a paraelectric $R\bar{3}c$ phase above Curie temperature (T$_c\approx$1483 K)[22–25].

Solid-solution of Lithium Niobate with Sodium Niobate i.e. Li$_x$Na$_{1-x}$NbO$_3$ (LNNx), shows a series of structural phase transitions with complex structures (supercells), as a function of composition (x). Many controversies have been reported in LNNx about the exact region of phase coexistence and Morphotropic Phase Boundary (MPB). Nitta et al. have reported a single phase for $x \leq 0.14$, and an additional lithium niobate like rhombohedral (R$_{Li}$) phase for LNN16[26]. Whereas, Jankowska et al. have observed an MPB like behaviour at $x \approx 0.04$, based on dielectric anomaly[27]. LNNx with x=0.12, has been identified as an MPB composition, based on dielectric properties, radial coupling coefficients, polarization, electromechanical response, and synchrotron X-ray diffraction measurements[28–31]. On the other hand, Mitra et al. have reported an MPB at LNN14 based on coexistence of R$_{Na}$ (low temperature ferroelectric *R3c* phase of NaNbO$_3$) and O$_{FE}$ (ferroelectric orthorhombic *Pmc*2$_1$) phases, using X-ray diffraction and Raman spectroscopy techniques, even though, they observed highest electrical properties for LNN12 having O$_{FE}$ phase[32]. On the other hand, Franco et al. didn't observe any MPB like behaviour in the region $0 \leq x \leq 0.12$[33]. Pozdnyakova et al. carried out a detailed structural phase transition studies in the regions $0 \leq x \leq 0.14$ and $0.95 \leq x \leq 1.00$ and reported a single phase (O$_{AFE}$) for $x \leq 0.0125$; two phases (O$_{AFE}$ & O$_{FE}$) for $0.015 \leq x \leq 0.02$; single phase (O$_{FE}$) for $0.0225 \leq x \leq 0.07$; two phases (O$_{FE}$ & R$_{Na}$) for $0.08 \leq x \leq 0.13$; single phase (O$_{FE}$) for $x = 0.14$ and again a single phase (R$_{Li}$) for $x \geq 0.95$[34]. Recently, Dixon et al. revised the composition-dependent phase diagram for $0.05 \leq x \leq 0.95$, using a combination of X-ray diffraction, neutron diffraction and, $^{23}$Na solid-state NMR spectroscopy technique[35–37]. They have reported a single phase (O$_{FE}$) for $x = 0.05$, two phases (O$_{FE}$ & R$_{Na}$) for $0.08 \leq x \leq 0.20$, two phases (R$_{Na}$ and R$_{Li}$) for $0.25 \leq x \leq 0.90$, and finally, a single phase for (R$_{Li}$) in the region $0.95 \leq x \leq 1.00$. They have also pointed out that the phase diagram depends on the synthesis conditions, such as annealing temperature, cooling rate etc.

In this letter, we have carried out detailed structural investigation using X-ray diffraction profiles for LNNx system, and strove to resolve the ambiguities present in the composition-dependent phase diagram of this system. Rietveld and Le-bail refinements revealed unambiguous evidence of three phase coexistence for region $0.15 \leq x \leq 0.80$. Composition-dependent dielectric and polarization measurements have shown anomalously high response for LNN20, similar to an MPB composition. These high responses near MPB compositions are related to the least energy barrier with flattened nature of free energy profile between the different phases, facilitating easy polarization rotation, similar to PZT[38].

The LNNx ($0 \leq x \leq 1.00$) ceramics were prepared using conventional solid-state reaction route (see the supplementary material for the details of synthesis). For phase transition studies, X-ray diffraction measurements were carried out

a)Electronic mail: stripathi.phy@itbhu.ac.in

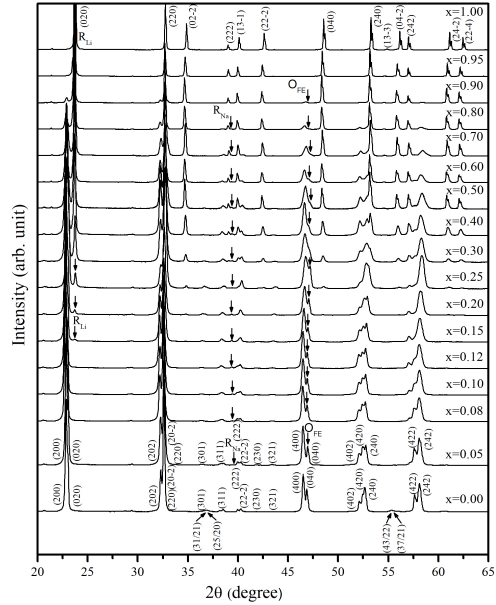

FIG. 1. X-ray diffraction pattern of the LNNx ceramics with varying Li content. The reflections are indexed with respect to doubled pseudocubic perovskite cell and unique reflections of different phases are shown by arrow marks.

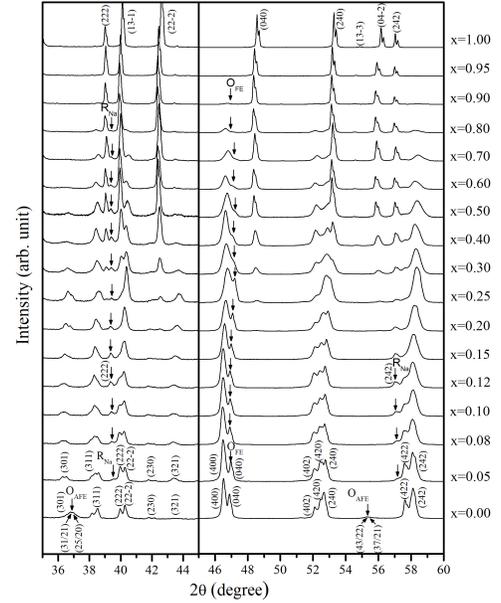

FIG. 2. Zoomed up view of portions of the X-ray diffraction patterns, showing evolution of $R_{Na}$ & $O_{FE}$ phases. The reflections are indexed with respect to doubled perovskite cell. Exclusive reflections of all the different phases are shown by arrow marks.

using a X-ray diffractometer (Rigaku MiniFlex 600), with Cu-$K_\alpha$ radiation. Diffraction data were collected at scan rate of 2°/minute and step interval of 0.02° in the $2\theta$ range of 20°-120°. The room temperature dielectric permittivity ($\varepsilon_r'$) and loss ($tan\delta$) measurements were performed on an impedance analyzer (HP4194A), at 100 kHz. The P-E hysteresis loops measurements of unpoled samples were carried out using Radiant Technology's Precision Material Analyzer Workstation based on virtual ground system at 10 Hz and an alternating voltage of about 70 kV/cm waveform.

X-ray diffraction studies confirm the formation of solid-solution for the entire composition range with a trace amount (<1%) of secondary phase i.e. $Na_{13}Nb_{35}O_{94}$ (Space group: $Pba2$)[39], due to alkali loss during sintering. Fig.1 shows composition-dependent evolution of powder X-ray diffraction patterns in the $2\theta$ range of 20°−65°. The reflections have been indexed in terms of doubled pseudocubic perovskite cell[31,40,41]. These reflections can be classified into two major categories viz; main perovskite reflections represented by all even (eee) integers, and superlattice reflections represented by one or more odd integer (s) such as (eeo), (ooe) or (ooo), and (or) fractional integers. Furthermore, (ooo) and (ooe) types of reflections represent out-of-phase and in- phase tilting of oxygen octahedra[42–44]. The X-ray diffraction patterns for $x = 0.00$, and $x = 1.00$, have been indexed with an orthorhombic anti-ferroelectric phase ($O_{AFE}$) having a compound tilt system $(a^-b^+a^-)_1^2(a^-b^-a^-)_2^3(a^-b^+a^-)_3^4$; (supercell∼ $\sqrt{2}a_p \times 4b_p \times \sqrt{2}c_p$) with quadrupling along [010] direction, and a rhombohedral ferroelectric phase ($R_{Li}$) with a simple tilt system $(a^-a^-a^-)$; (supercell≡ $\sqrt{2}a_p \times \sqrt{2}b_p \times 2\sqrt{3}c_p$) respectively, consistent with the literature[19,24,42,43,45]. For a small amount of $Li^+$ substitution ($x = 0.05$), the structure of LNNx transforms to a ferroelectric $O_{FE}$ phase as shown in Fig.1, and Fig.2 (for comparison, indexing has been done in $cab$ setting of space group no. 26). Note that, for indexing the X-ray diffraction profiles, we have tried all the isotropy subgroups of $Pm\bar{3}m$, resulting via condensation of $R_4^+$ (q=1/2,1/2,1/2) and $M_3^+$ (q=1/2,1/2,0) soft phonon modes leading to the simple tilt systems originally given by Glazer[42,43,46]. We have also tried the isotropy subgroups of $Pm\bar{3}m$, for the irreducible representations (IRs) viz; $GM_4^-$ (q=0,0,0), $R_4^+$ (q=1/2,1/2,1/2), and $M_3^+$ (q=1/2,1/2,0), leading to structures corresponding to the Glazer's simple tilt system with ferroelectric displacements[42,43,46]. We found that the structure could be indexed with ferroelectric $Pmc2_1$ space group (supercell ∼ $2a_p \times \sqrt{2}b_p \times \sqrt{2}c_p$ having tilt system $a_+^-a_+^-c_0^+$ in the modified Glazer's notation[46]) using Le-bail refinements and crystallographic open database[47,48]. The $O_{AFE}$ to $O_{FE}$ transition is supported by the disappearance of superlattice reflections corresponding to the freezing of $T_4$ (q=1/2,1/4,1/2) phonon mode of $Pm\bar{3}m$ space group, related with the quadrupling of $b_p$ unit-cell parameter. Example of such reflections are (3 1/2 1), (2 5/2 0) at $2\theta \approx 37°$ and (4 3/2 2), (3 7/2 1) at $2\theta \approx 55.5°$ (Fig.1 and Fig.2). The $O_{AFE}$ to $O_{FE}$ transition is also reported by Pozdnyakova et al.[34] and Yuzyuk et al.[31]. Furthermore, the most intense reflection (200) and the exclusive reflection (040) of $O_{FE}$ phase at $2\theta = 22.9°$ and $47.0°$ respectively, disappears at $x = 0.95$, suggesting a single $R_{Li}$ ($a_+^- a_+^- a_+^-$ tilt system in the modified Glazer's notation[46]) phase, for $x \geq 0.95$, similar to the reports by Pardo et al. and Pozdnyakova et al[34,49]. On the other hand, the most intense peak of LiNbO$_3$ ($R_{Li}$ phase) corresponds to (020) reflection at $2\theta \approx 23.6°$ (Fig.1), survives till $x = 0.15$, asserting the existence of $R_{Li}$ phase in the region $0.15 \leq x \leq 1.00$. The presence of $R_{Li}$ phase in this region is also supported by Nitta et al. and Pardo et al[26,49]. On the other hand, Mishra et al. found a small phase fraction ($\approx 2\%$) of $R_{Li}$ phase even at $x = 0.12$[50]. The evolution of (ooo) reflection (1 3 -1) of LN at $2\theta \approx 40°$ corresponding to $R_{Li}$ phase, survives till $x = 0.20$, however for $x = 0.15$, it seems to have merged with background due to the small phase fraction of $R_{Li}$. Moreover, on close inspection of diffraction pattern of LNN5, we find that the reflection at $2\theta \approx 39.4°$ can not be indexed with $O_{FE}$ or $O_{AFE}$ phase. Presence of such reflections have been previously reported by Yuzyuk et al. and Mishra et al. in the vicinity of LNN12. They claimed that, this phase was similar to the low temperature rhombohedral $R_{Na}$ ($a_+^- a_+^- a_+^-$, in modified Glazer's notation[46]) phase of NaNbO$_3$[31,50]. The evolution of $R_{Na}$ phase is shown by its exclusive reflection (222) at $2\theta = 39.4°$, as clearly depicted in Fig. 2. The existence of reflection corresponding to $R_{Na}$ phase for $0.05 \leq x \leq 0.80$ clearly shows that this composition range can not be indexed with $O_{FE}$ phase or two phase model ($O_{FE} + R_{Li}$). Moreover, the possibility of indexing these structures with two phase model ($R_{Na} + R_{Li}$), as reported by Peel et al.[35] can be ruled out from the evolution of $O_{FE}$ reflection at $2\theta = 46.82°$ (Fig.2). The evolution of this exclusive reflection confirms the presence of $O_{FE}$ phase from $x = 0.05$ to $x = 0.80$. Therefore, for compositions $0.15 \leq x \leq 0.80$, we have undoubtedly three coexisting ferroelectric phases, $O_{FE}$, $R_{Na}$ and $R_{Li}$. Also, the relative intensity of the $R_{Li}$ phase most intense and exclusive reflection in LNN20 is $\approx 0.0369$ or $3.69\%$ at $2\theta = 23.72°$ (marked by the arrow in Fig.1). Similarly, the relative intensity of the exclusive reflection corresponding to $R_{Na}$ phase is $\approx 0.0074$ or $0.74\%$ in LNN20 at $2\theta = 39.36°$ (marked by arrow in Fig.2).

To further ascertain the composition-dependent phase diagram, we have performed Rietveld refinements of X-

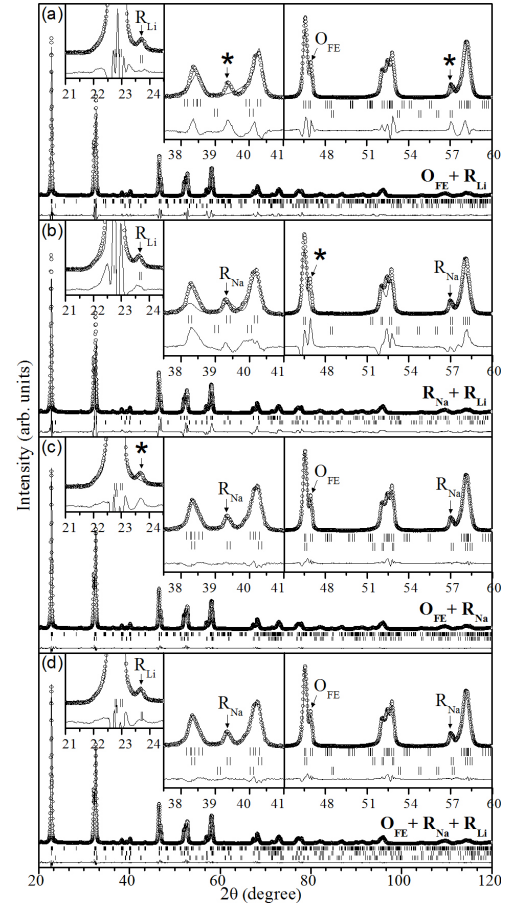

FIG. 3. Rietveld refinements of LNN15 with various structural models (a) $O_{FE}+R_{Li}$, (b) $R_{Na}+R_{Li}$, (c) $O_{FE}+R_{Na}$, and (d) $O_{FE}+R_{Na}+R_{Li}$. Open circles represent the observed diffraction pattern, continuous profile curve is calculated model, vertical bars are Braggs' positions and horizontal line at the bottom is the difference profile between observed and calculated patterns. Reflections with $\star$ mark could not be indexed with the specified phases.

TABLE I. Composition-dependent phases of LNNx.

| Composition | Phases |
|---|---|
| $x = 0.00$ | $O_{AFE}$ |
| $0.05 \leq x \leq 0.12$ | $O_{FE} + R_{Na}$ |
| $0.15 \leq x \leq 0.80$ | $O_{FE} + R_{Na} + R_{Li}$ |
| $x = 0.90$ | $O_{FE} + R_{Li}$ |
| $0.95 \leq x \leq 1.00$ | $R_{Li}$ |

ray diffraction data, using a structure refinement program "FULLPROF"[48]. The Pseudo-Voigt function was chosen for peak profile fitting in the refinements. The Background was described in terms of linear interpolation between a set background points with refinable heights. The lattice parameters, zero correction, half width parameters (U, V and W), atomic coordinates and thermal parameters were refined to get least square (chi square) fit between the observed and simulated powder diffraction pattern. In order to confirm three phase coexistence, we have carried out Rietveld refinements of LNN15, using various models as shown in Fig.3(a)-(d). Fig.3(a) shows fitted profile of LNN15 using two phase model ($O_{FE}$ + $R_{Li}$), similar to the reports by Nitta et al. and Pardo et al[26,49]. However, we could not index the reflections present at $2\theta = 39.4°$ and $57.1°$ (marked with ⋆), by using this two phase model. Similarly, Fig.3(b) shows fitted profile with two phase model ($R_{Na}$ + $R_{Li}$), which results in even worse fit, as is clearly evident from the poor profile fitting and the absence of peak in the simulated pattern corresponding to $O_{FE}$ phase at $2\theta = 46.8°$. Furthermore, we have refined the profile of LNN15 with the widely reported two phase model ($O_{FE}$ + $R_{Na}$)[32,35,37,50]. This model could not index the reflection at $2\theta = 23.7°$ corresponding to $R_{Li}$ phase, as shown by ⋆ mark in the inset of Fig.3(c). Therefore, it is clear that none of the two phase model could index all the reflections present in LNN15. Fig.3(d) shows the fitted profile with three phase model ($O_{FE}$ + $R_{Na}$ + $R_{Li}$), giving the best fit with lowest agreement factors. Although, for compositions $0.15 \leq x \leq 0.80$, the best fit has been achieved by three phase ($O_{FE}$, $R_{Na}$ and $R_{Li}$) model, we observed an extra reflection with very low intensity in the vicinity of ($2\theta = 36.4°$) for compositions $0.15 \leq x \leq 0.25$. This reflection may be due to the presence of an incommensurate modulation or long-range ordering which requires high-resolution neutron/synchrotron diffraction data to fix the structure[37]. The revised phase diagram as confirmed from Rietveld and Le-bail refinements, can be seen from Table I.

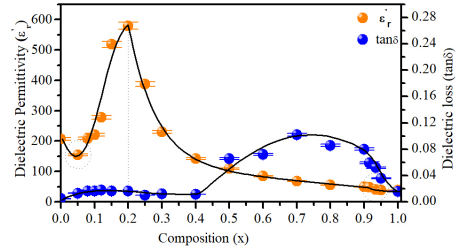

FIG. 4. Composition-dependent dielectric permittivity and loss measured at 100 kHz.

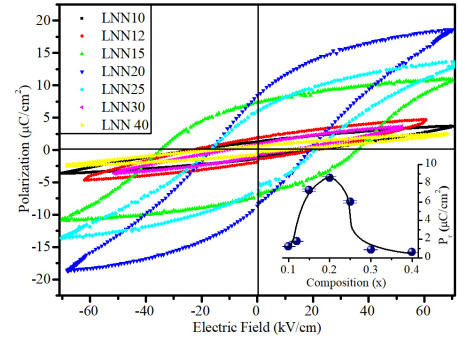

FIG. 5. Room temperature P-E hysteresis loop and Remnant polarization ($P_r$) for LNNx ceramics under field of 70kV/cm and frequency 10 Hz.

The variation of room temperature dielectric permittivity ($\varepsilon'_r$) and loss ($tan\delta$) as a function of composition is shown in Fig.4. The $\varepsilon'_r$ values of undoped NN and LN are 207 and 38 respectively. These values are close to the reported values in literature[51–55]. On $Li^+$ substitution, $\varepsilon'_r$ first decreases for LNN5, and then starts increasing. The anomaly at LNN5 can be related to the onset of $R_{Na}$ phase. The dielectric permittivity shows surprisingly high value of 518 for LNN15 and attains a maximum value of 579 for LNN20, and then again starts decreasing. The peaking behaviour of dielectric response in the vicinity of LNN20 can be attributed to the MPB like behaviour around this composition[38]. Henson et al. have also reported an anomaly at LNN12, even though, $\varepsilon'_r$ values were less than 200 for all the compositions $0.02 \leq x \leq 0.15$[28]. For $x > 0.20$, $\varepsilon'_r$ decreases continuously on further increasing

Li content. We also observe a weak anomaly around LNN95, which is due to the disappearance of $O_{FE}$ phase. We could not find a single report for LNNx, showing similar anomalies in dielectric permittivity with such high dielectric response. Furthermore, the dielectric loss ($tan\delta$), exhibit small values ($< 0.018$) for compositions $x \leq 0.40$, which is of the same order as reported by Mitra et al.[32].

We have performed the room temperature PE hysteresis loop measurement for various compositions $0.10 \leq x \leq 0.40$ in the vicinity of LNN20 (Fig.5). The variation of remnant polarization ($P_r$) for $0.10 \leq x \leq 0.40$, clearly shows the presence of high $P_r$ values in LNNx for $0.15 \leq x \leq 0.25$ (Fig.5). LNN15 and LNN20, both have comparable $P_r$ (7.26 & 8.60 $\mu C/cm^2$ respectively), but high value of coercive field ($E_c$ = 35.5 kV/cm) makes LNN15 inappropriate for applications. The high $E_c$ in LNN15 is linked with lower experimental density of LNN15 (4.463 gm/cc) w.r.t. LNN20 (4.528 gm/cc)[56]. On the other hand, high $\varepsilon'_r$ (579) and $P_r$ (8.60 $\mu C/cm^2$), along with low $E_c$ (17.49 kV/cm) and low loss ($tan\delta$ = 0.016) makes LNN20, a suitable candidate for ferroelectric devices. The polarization values are in close agreement with Mitra et al.[32]. Note that, LNNx ceramics become highly conductive at high voltage of around 70kV/cm, and therefore, saturation could not be achieved for many compositions and hence, $P_r$ values are compared at a field of 70kV/cm.

In conclusion, composition-dependent X-ray diffraction analysis reveals several phase transitions in LNNx ceramics, along with three coexisting ferroelectric phases stable in region $0.15 \leq x \leq 0.80$ unlike Dixon et al. reports, where they have reported a coexistence of $O_{FE}$ and $R_{Na}$ phases for $0.15 \leq x \leq 0.20$, and $R_{Na}$ and $R_{Li}$ phases for $0.20 < x \leq 0.90$. Room temperature dielectric and polarization measurements have shown an anomaly at LNN20 and is attributed to the MPB-like behaviour near this composition, which is further confirmed by the temperature-dependent X-ray diffraction studies of LNN20 showing the existence of all the three phases up to 423K (see Fig. S1 in supplementary materials). The MPB-like behaviour is related with the least energy barrier between the different phases, facilitating easy polarization rotation, similar to PZT system, where the two phases (P4mm+R3m) coexist over the composition range $0.451 < Ti(x) < 0.488$, but the highest properties has been observed for $Ti(x) = 0.48$[38,57–59]. These results suggest an MPB like behaviour for LNNx at $x \approx 0.20$, against widely reported MPB at $x = 0.12$.

**SUPPLEMENTARY MATERIAL**

See supplementary materials for the materials synthesis procedure, temperature-dependent X-ray diffraction analysis, and the variation of dielectric permittivity and loss with composition (x) for LNNx ceramics.

The data that supports the findings of this study are available within the article and its supplementary material.

[1] J. Gordon, A. Taylor, and P. Bennett, British journal of clinical pharmacology **53**, 451 (2002).
[2] Official Journal L , 0019 (2003).
[3] S. O. Leontsev and R. E. Eitel, Science and Technology of Advanced Materials **11**, 044302 (2010).
[4] E. Cross, Nature **432**, 24 (2004).
[5] Y. Saito, H. Takao, T. Tani, T. Nonoyama, K. Takatori, T. Homma, T. Nagaya, and M. Nakamura, Nature **432**, 84 (2004).
[6] W. P. Mason, The Journal of the Acoustical Society of America **70**, 1561 (1981).
[7] W. Zhu, B. Akkopru-Akgun, and S. Trolier-McKinstry, Applied Physics Letters **111**, 212903 (2017).
[8] Y. Chang, S. F. Poterala, Z. Yang, S. Trolier-McKinstry, and G. L. Messing, Applied Physics Letters **95**, 232905 (2009).
[9] Y. Yang, Y. Ji, M. Fang, Z. Zhou, L. Zhang, and X. Ren, Physical review letters **123**, 137601 (2019).
[10] M. E. Lines and A. M. Glass, *Principles and applications of ferroelectrics and related materials* (Oxford university press, 2001).
[11] G. Maciel, N. Rakov, C. B. de Araujo, A. Lipovskii, and D. Tagantsev, Applied Physics Letters **79**, 584 (2001).
[12] E. Falcão-Filho, C. Bosco, G. Maciel, L. Acioli, C. B. de Araújo, A. Lipovskii, and D. Tagantsev, Physical Review B **69**, 134204 (2004).
[13] E. Valdez, C. B. de Araújo, and A. Lipovskii, Applied physics letters **89**, 031901 (2006).
[14] A. A. Ballman, Journal of the American Ceramic Society **48**, 112 (1965).
[15] K. Uchino, *Ferroelectric Devices 2nd Edition* (CRC press, 2009).
[16] M. Fontana, S. Mignoni, R. Hammoum, and P. Bourson, in *International Union of Crystallography (IUCr)* (2011).
[17] Y. Xu, *Ferroelectric materials and their applications* (Elsevier, 2013).
[18] J. He, C. Franchini, and J. M. Rondinelli, Chemistry of Materials **28**, 25 (2015).
[19] H. D. Megaw, Ferroelectrics **7**, 87 (1974).
[20] S. Mishra, R. Mittal, V. Y. Pomjakushin, and S. Chaplot, Physical Review B **83**, 134105 (2011).
[21] L. Jiang, D. Mitchell, W. Dmowski, and T. Egami, Physical Review B **88**, 014105 (2013).
[22] B. Matthias and J. Remeika, Physical Review **76**, 1886 (1949).
[23] B. T. Matthias, Science **113**, 591 (1951).
[24] H. D. Megaw, Acta Crystallographica **7**, 187 (1954).
[25] H. D. Megaw, Acta Crystallographica Section A: Crystal Physics, Diffraction, Theoretical and General Crystallography **24**, 583 (1968).
[26] T. Nitta, Journal of the American Ceramic Society **51**, 623 (1968).
[27] I. Jankowska, K. Krzywanek, C. Kus, and W. S. Ptak, Ferroelectrics **127**, 83 (1992).
[28] R. Henson, R. Zeyfang, and K. Kiehl, Journal of the American Ceramic Society **60**, 15 (1977).
[29] B. Hardiman, R. Henson, C. Reeves, and R. Zeyfang, Ferroelectrics **12**, 157 (1976).
[30] P. Zhang, W. Zhong, H. Chen, F. Chen, and Y. Song, Japanese Journal of Applied Physics **24**, 998 (1985).
[31] Y. I. Yuzyuk, E. Gagarina, P. Simon, L. Reznitchenko, L. Hennet, and D. Thiaudiere, Physical Review B **69**, 144105 (2004).
[32] S. Mitra, A. Kulkarni, and O. Prakash, Journal of Applied Physics **114**, 064106 (2013).
[33] R. C. Franco, E. R. Camargo, M. A. Nobre, E. R. Leite, E. Longo, and J. A. Varela, Ceramics international **25**, 455 (1999).
[34] I. Pozdnyakova, A. Navrotsky, L. Shilkina, and L. Reznitchenko, Journal of the American Ceramic Society **85**, 379 (2002).
[35] M. D. Peel, S. E. Ashbrook, and P. Lightfoot, Inorganic chemistry **52**, 8872 (2013).
[36] C. A. Dixon, J. A. McNulty, S. Huband, P. A. Thomas, and P. Lightfoot, IUCrJ **4**, 215 (2017).
[37] C. A. Dixon and P. Lightfoot, Physical Review B **97**, 224105 (2018).
[38] M. Soares, A. Senos, and P. Mantas, Journal of the European Ceramic Society **20**, 321 (2000).
[39] D. Craig and N. Stephenson, Journal of Solid State Chemistry **3**, 89 (1971).
[40] H. D. Megaw and C. Darlington, Acta Crystallographica Section A: Crystal Physics, Diffraction, Theoretical and General Crystallography **31**, 161 (1975).
[41] S. Tripathi, D. Pandey, S. K. Mishra, and P. Krishna, Physical Review B **77**, 052104 (2008).
[42] A. M. Glazer, Acta Crystallographica Section B: Structural Crystallography and Crystal Chemistry **28**, 3384 (1972).
[43] A. M. Glazer, Acta Crystallographica Section A: Crystal Physics, Diffraction, Theoretical and General Crystallography **31**, 756 (1975).
[44] C. J. Howard and H. T. Stokes, Acta Crystallographica Section B: Structural Science **54**, 782 (1998).
[45] C. Darlington and K. Knight, Physica B: Condensed Matter **266**, 368 (1999).
[46] H. T. Stokes, E. H. Kisi, D. M. Hatch, and C. J. Howard, Acta Crystallographica Section B: Structural Science **58**, 934 (2002).
[47] A. Merkys, A. Vaitkus, J. Butkus, M. Okulič-Kazarinas, V. Kairys, and S. Gražulis, Journal of applied crystallography **49**, 292 (2016).
[48] J. Rodriguez-Carvajal and A. FULLPROF, Laboratory Leon Brillouin (CEA-CNRS), France (2011).
[49] L. Pardo, P. Duran-Martin, J. Mercurio, L. Nibou, and B. Jimenez, Journal of Physics and Chemistry of Solids **58**, 1335 (1997).
[50] S. Mishra, P. Krishna, A. Shinde, V. Jayakrishnan, R. Mittal, P. Sastry, and S. Chaplot, Journal of Applied Physics **118**, 094101 (2015).
[51] B. Matthias and J. Remeika, Physical Review **82**, 727 (1951).
[52] V. Akhnazarova, O. Y. Kravchenko, L. Shilkina, and L. Reznichenko, Inorganic Materials **47**, 554 (2011).
[53] S. Bhatt and B. Semwal, Solid state ionics **23**, 77 (1987).
[54] R. Smith and F. Welsh, Journal of applied physics **42**, 2219 (1971).
[55] C. R. Cena, A. K. Behera, and B. Behera, Journal of Advanced Ceramics **5**, 84 (2016).
[56] P. Bharathi and K. Varma, Journal of Applied Physics **116**, 164107 (2014).
[57] Z. Wu and R. E. Cohen, Physical review letters **95**, 037601 (2005).
[58] D. Damjanovic, Applied Physics Letters **97**, 062906 (2010).
[59] M. Porta and T. Lookman, Physical Review B **83**, 174108 (2011).

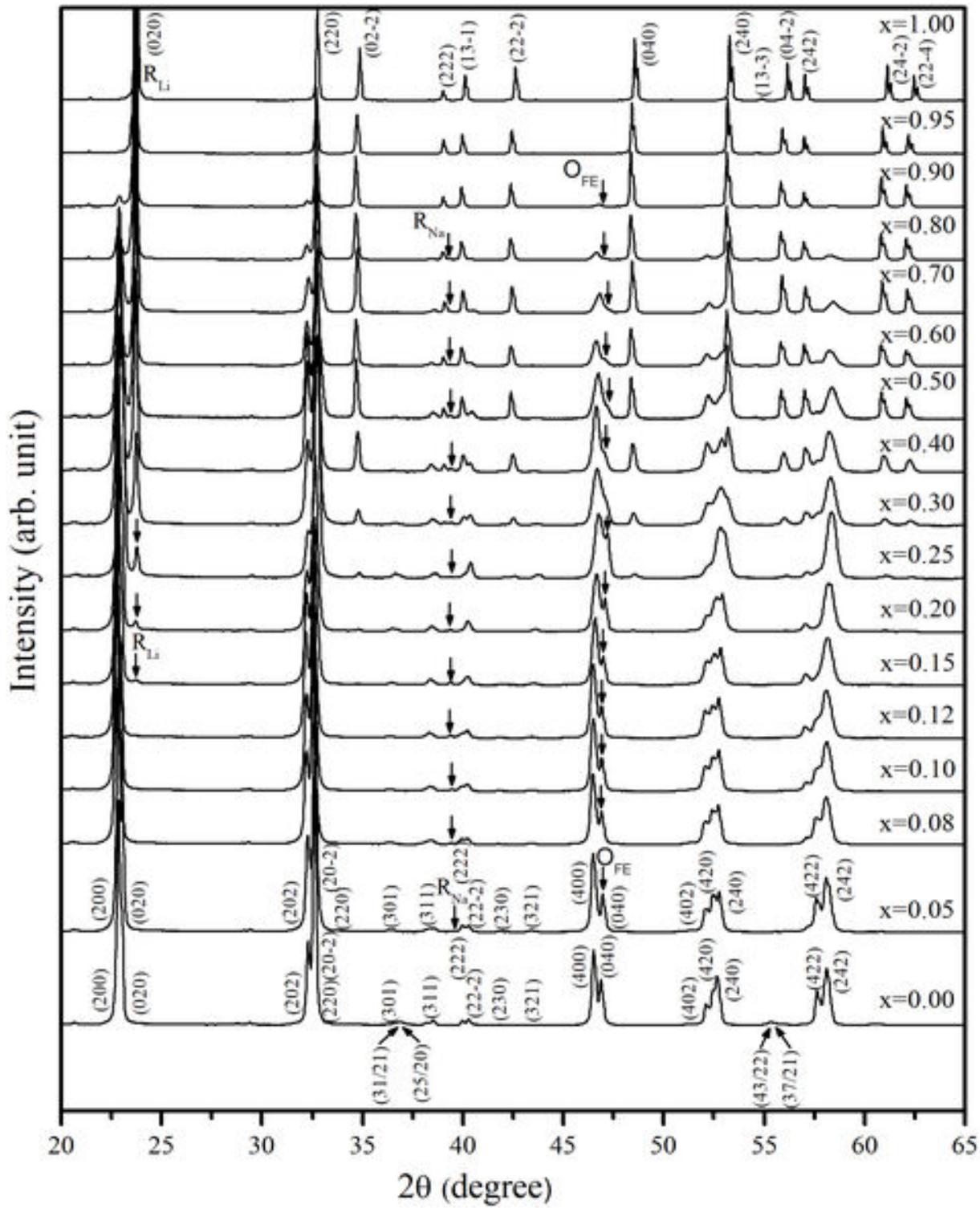

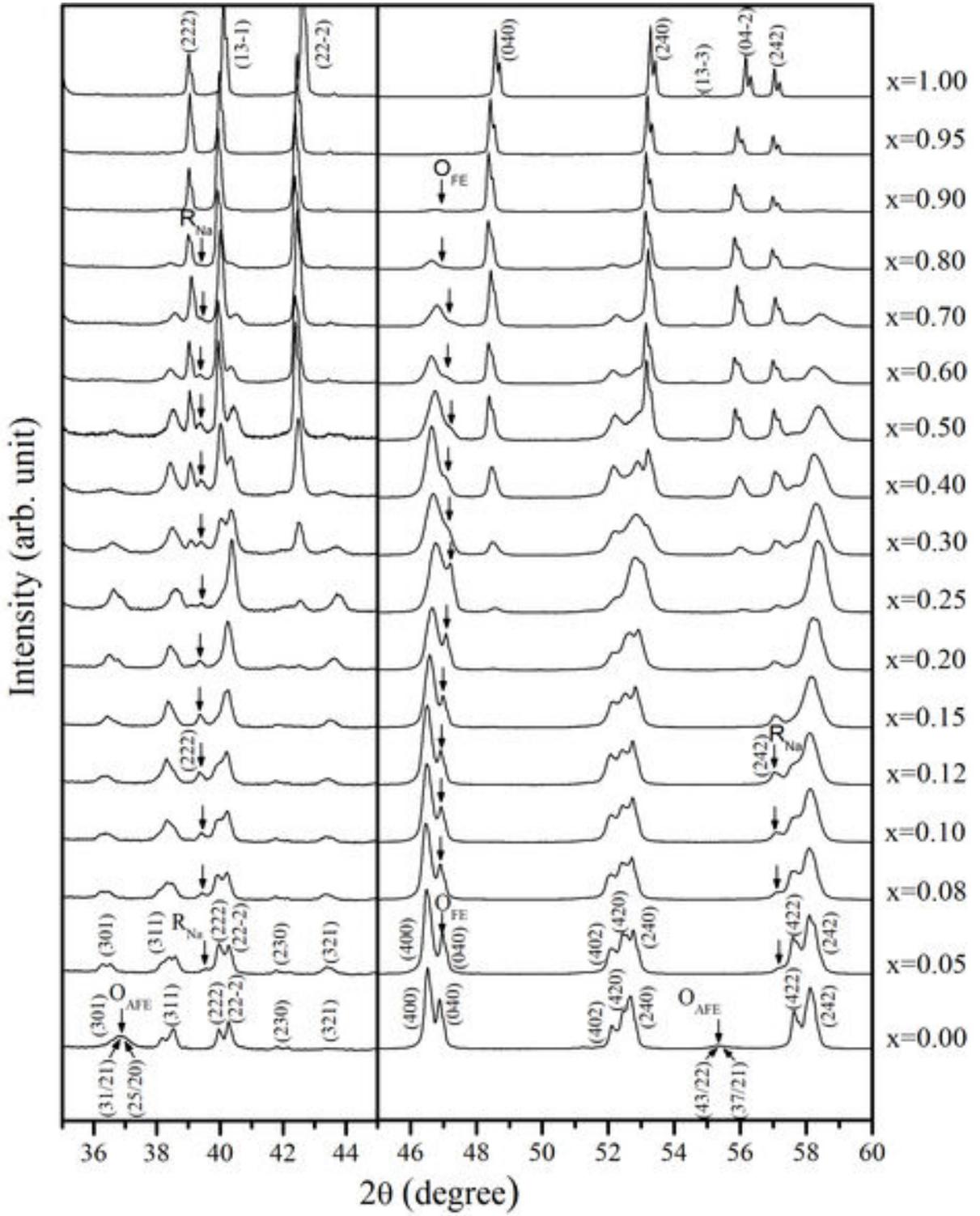

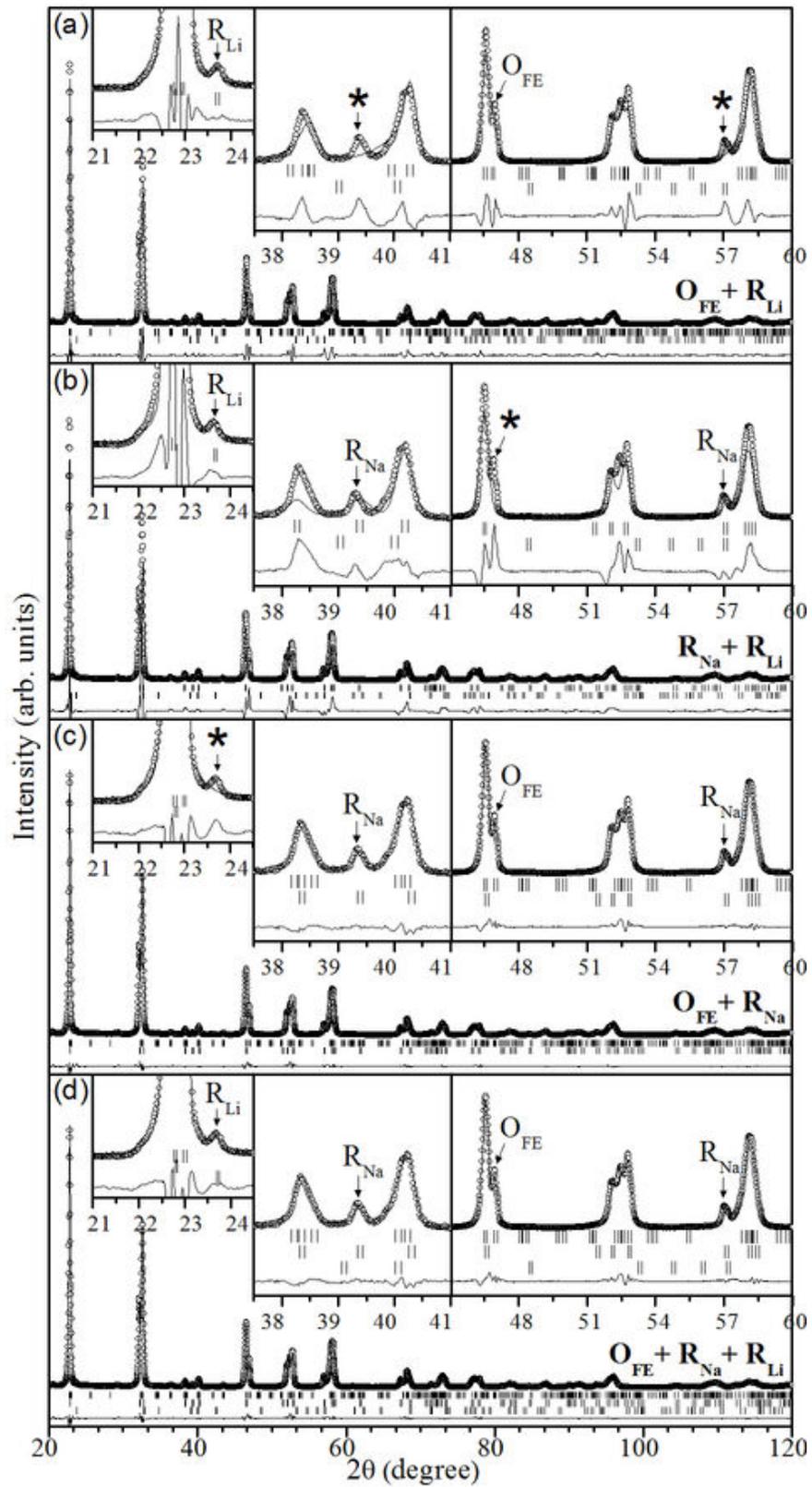

<: />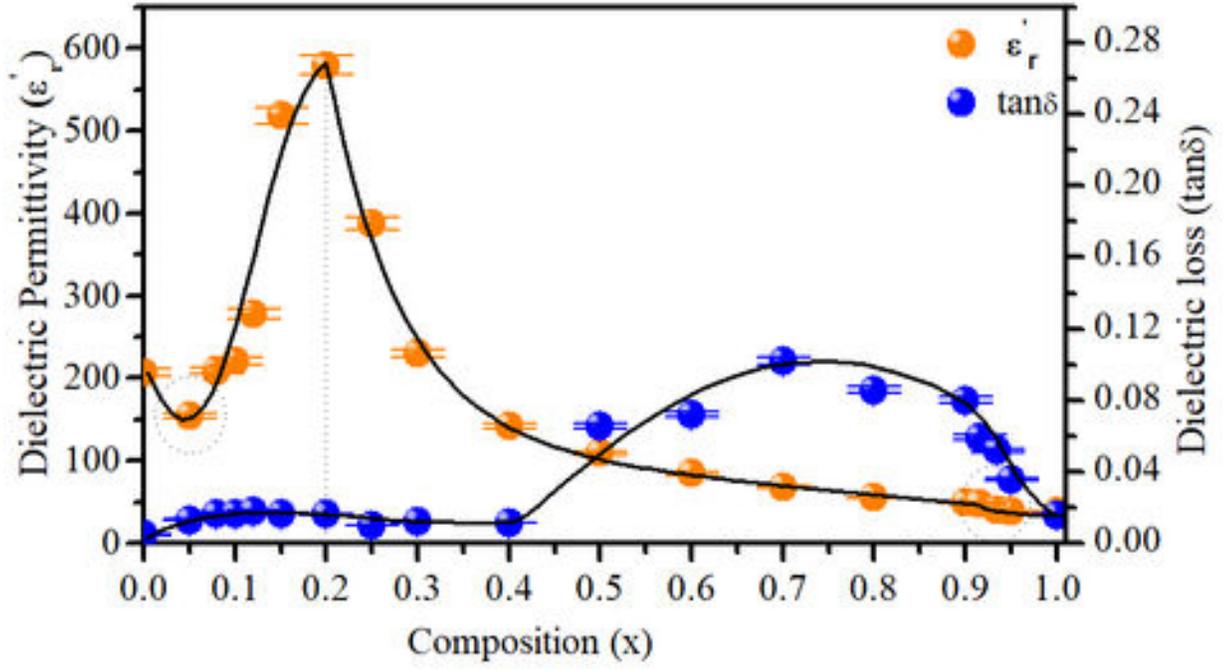

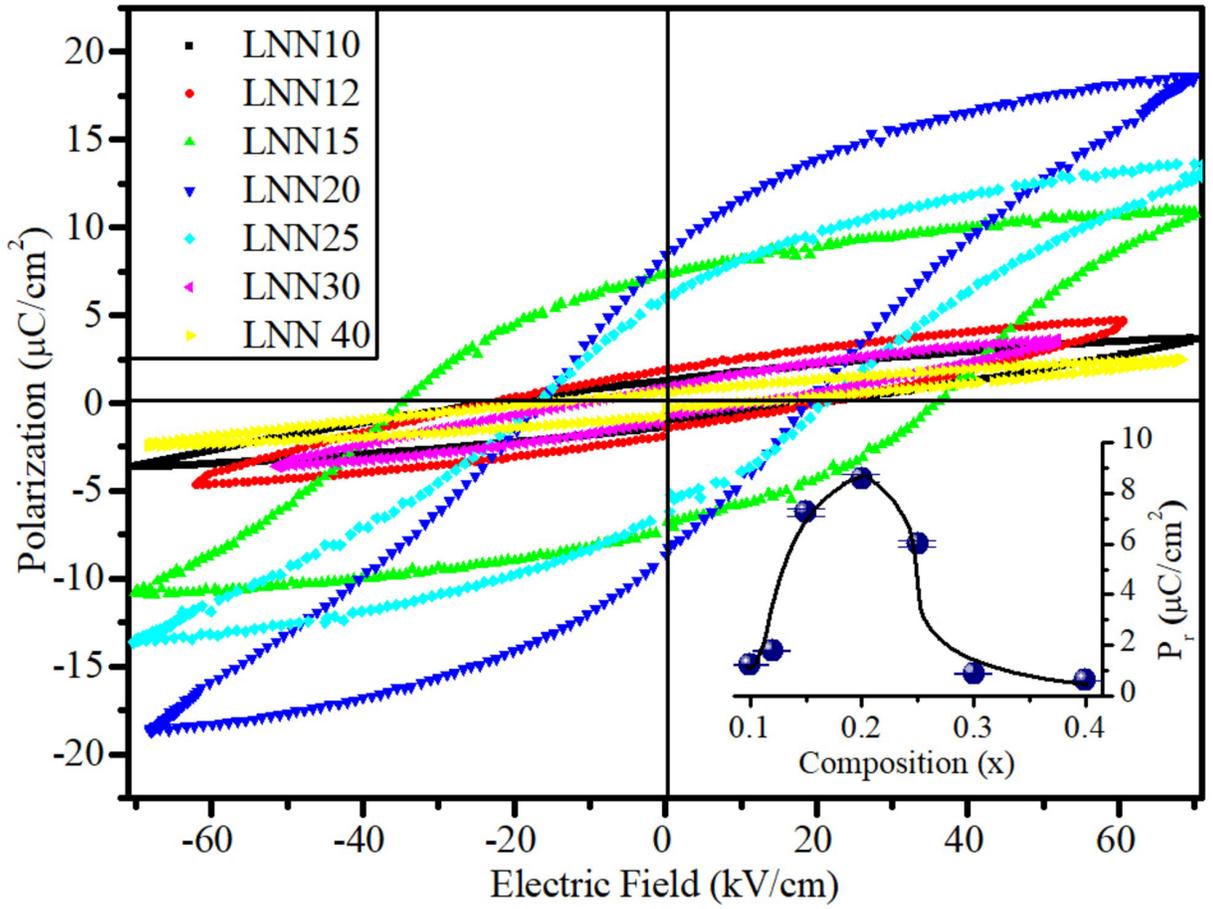